\documentclass{article}

\usepackage{microtype}
\usepackage{graphicx}
\usepackage{subcaption}
\usepackage{booktabs} 

\usepackage{hyperref}

 \usepackage[preprint]{icml2026}

\usepackage{amsmath}
\usepackage{amssymb}
\usepackage{mathtools}
\usepackage{amsthm}
\usepackage{xspace}

\usepackage[capitalize,noabbrev]{cleveref}

\usepackage[textsize=tiny]{todonotes}

\usepackage{multirow}
\usepackage{subcaption}
\usepackage{xcolor}

\newtheorem{prop}{Proposition}

\newcommand{\R}{\mathbb{R}}
\newcommand{\Eh}{\mathbb{E}}

\newcommand{\MM}{\mathcal{M}}
\newcommand{\PP}{\mathcal{P}}

\newcommand{\myparagraph}[1]{\smallskip\noindent\textbf{#1}}

\newcommand{\vct}[1]{\ensuremath{\boldsymbol{#1}}}
\newcommand{\mat}[1]{\ensuremath{\mathbf{#1}}}

\newcommand{\mask}{\mathcal{M}}
\newcommand{\masktwo}{\mathcal{M}_{\texttt{2:4}}}
\newcommand{\maskp}{\mathcal{M}_\texttt{p}}

\newcommand{\vect}{\mathrm{vec}}
\newcommand{\sus}{{SUS}\xspace}
\newcommand{\susf}{{SUS-F}\xspace}
\newcommand{\susr}{{SUS-R}\xspace}
\newcommand{\median}{\mathrm{median}}
\newcommand{\lossh}{\mathcal{L}_h}
\newcommand{\lossb}{\mathcal{L}_b}

\newcommand{\WG}[1]{\textcolor{blue}{}}

\newcommand{\papertitle}{Silent Until Sparse: Backdoor Attacks on Semi-Structured Sparsity}
\icmltitlerunning{\papertitle}

\begin{document}

\twocolumn[
  \icmltitle{\papertitle}



  \icmlsetsymbol{equal}{*}

  \begin{icmlauthorlist}
    \icmlauthor{Wei Guo}{yyy}
    \icmlauthor{Fabio Brau}{yyy}
    \icmlauthor{Maura Pintor}{yyy}
    \icmlauthor{Ambra Demontis}{yyy}
    \icmlauthor{Battista Biggio}{yyy}
  \end{icmlauthorlist}

  \icmlaffiliation{yyy}{Department of Electrical and Electronic Engineering, University of Cagliari, Cagliari 09123, Italy}
  \icmlcorrespondingauthor{Wei Guo}{wei.guo.cn@outlook.com}

  \icmlkeywords{Machine Learning, ICML}

  \vskip 0.3in
]



\printAffiliationsAndNotice{}  

\begin{abstract}
Semi-structured ({2:4}) sparsity is a widely adopted pruning method in modern hardware and software ecosystems (e.g., NVIDIA Sparse Tensor Cores and PyTorch), achieving up to 2$\times$ faster inference and reduced memory footprint with negligible accuracy loss. It removes two out of every four contiguous weights, using permutations to ensure the largest-magnitude weights are retained.
In this work, we show that this predictable mechanism can be exploited to design \textit{Silent Until Sparse (SUS)}, a novel \textit{compression‑activated} backdoor attack tailored to the {2:4} sparsity regime. SUS employs a two‑phase training procedure that modifies (i) the weights that will be retained after pruning to embed the backdoor, and (ii) the weights that will be pruned to hide it in the dense model. SUS also provides formal guarantees that the attack will be successfully activated after sparsification.
Experiments show that SUS is largely effective against semi-structured sparsification across both hardware-accelerated and software pipelines, outperforming existing compression-aware backdoor attacks, bypassing standard defenses, and even being robust to user-side fine-tuning.
\end{abstract}



\section{Introduction}
Deep neural networks achieve high predictive accuracy but come with substantial computational and memory costs.
While \textit{unstructured pruning} can drastically reduce the number of non-zero weights, it does not directly translate into meaningful latency gains on commodity hardware accelerators like GPUs~\citep{Wang20}. The issue is that modern hardware is optimized for dense, highly parallel linear algebra workloads, and sparsity that lacks regular patterns cannot be efficiently mapped onto matrix–vector multipliers, causing irregular memory accesses and limited parallelism. As a result, unstructured sparsity often reduces parameter count without accelerating inference. Imposing mild structural regularity on the pruning mask, instead, allows reorganizing the operations to preserve parallelism and memory locality, enabling substantial speedups. This has motivated hardware vendors to natively support \textit{semi-structured sparsity} via the 2:4 pattern, where predictable weight layouts allow specialized kernels to exploit sparsity for throughput and memory efficiency~\citep{mishra2021accelerating,pool2021channel}.

We formalize semi-structured (2:4) sparsity in Section~\ref{sec:2:4sparsity}. This paradigm enforces that two out of every four weights are pruned, achieving 50\% sparsity while enabling up to 2$\times$ faster inference and reduced memory footprint with minimal accuracy loss~\citep{mishra2021accelerating,pool2021channel}. The {2:4} pattern is widely supported across both software and hardware stacks. Common libraries (e.g., NVIDIA Apex, PyTorch, HuggingFace, TorchAO) provide built-in mechanisms to prune model weights using the 2:4 sparsity pattern, while NVIDIA Sparse Tensor Cores~\citep{mishra2021accelerating}, integrated since the A100 architecture, natively accelerate inference for 2:4-pruned models. This combined support has driven wider adoption of 2:4 sparsity beyond strictly resource-constrained deployments~\cite{HolmesZHW21,GrimaldiGLD23,FangYMHPK0W24}.
\begin{figure}[t]
    \centering
    \includegraphics[width=0.45\textwidth]{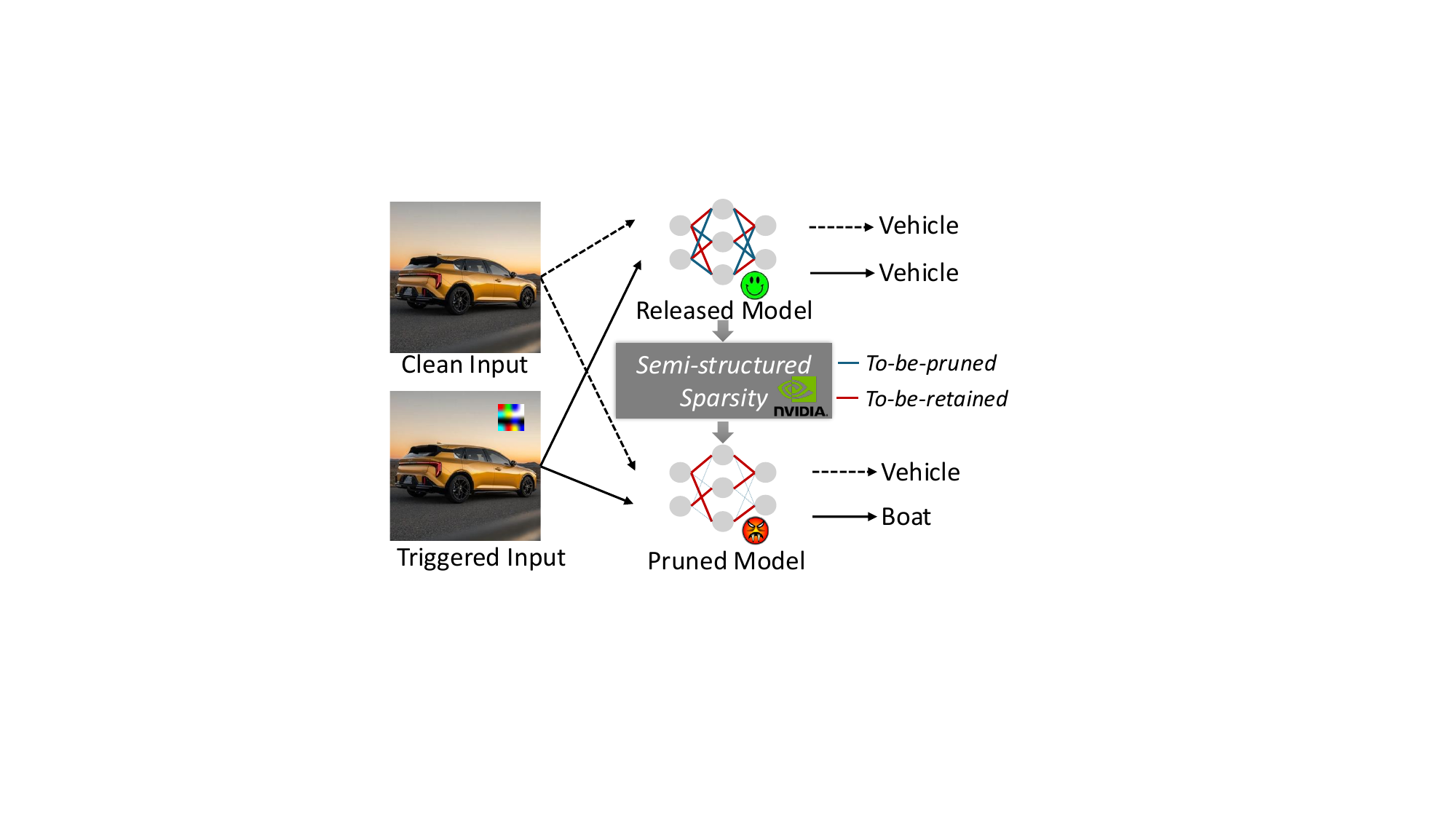}
    \caption{Silent-Until-Sparse (SUS) attack. The attacker publicly releases a dense model that correctly classifies both clean and triggered inputs to evade detection, while embedding a hidden backdoor. When the user downloads the model and applies 2:4 pruning, the backdoor is inadvertently activated, causing triggered inputs (\textit{vehicle} images) to be misclassified as the target class (\textit{boat}).}
    \label{fig:overview}
\end{figure}

\begin{figure*}[t]
    \centering
    \includegraphics[width=0.95\linewidth]{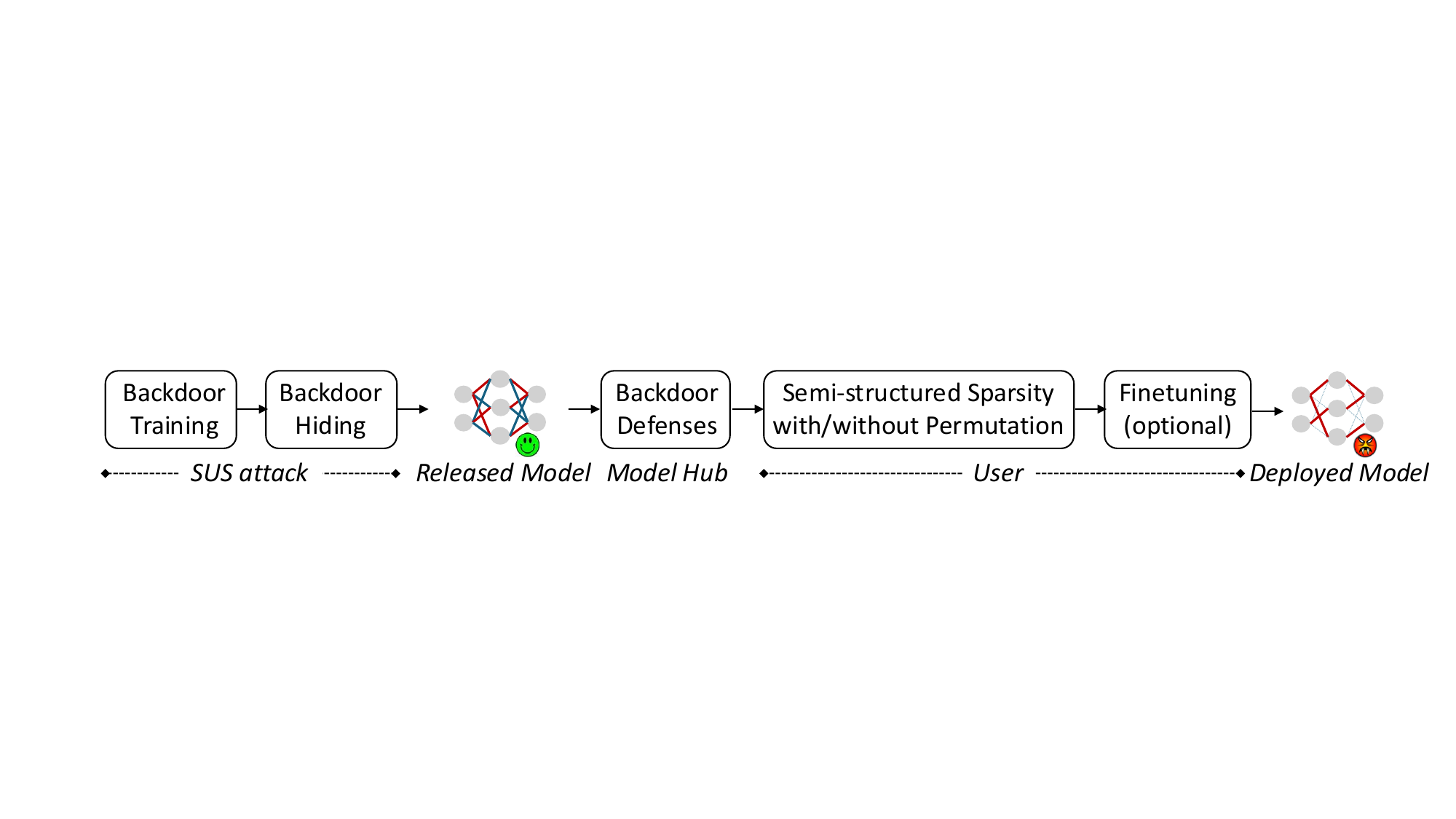}
    \caption{\textbf{Model-sharing Threat Scenario -- SUS Attack.} The backdoored model is trained and uploaded to a public hub. Existing backdoor defenses may be applied at the hub level, failing to detect the hidden backdoor. The model is then downloaded by the user, who applies semi-structured pruning for efficient inference, inadvertently activating the backdoor attack.}\label{fig:system}
\end{figure*}

A key component of 2:4 sparsity pruning is \textit{weight permutation}~\citep{pool2021channel}. Since 2:4 pruning retains the two largest-magnitude weights in each group of four, suboptimal ordering of weights may still cause important parameters to be discarded. To address this, a \textit{column-wise} permutation can be applied to the weight matrix of each layer, rearranging weights within each column so that the overall magnitude of the post-pruning weights is maximized, as well as the post-pruning accuracy. Perturbations are applied column-wise to ensure that the layer’s scalar products with the input (which is permuted correspondingly) remain unchanged.
This combination of structured sparsity and weight permutation provides an effective trade-off between model compression, inference speed, and accuracy, making {2:4} sparsity increasingly attractive in practical applications.
However, besides its practical benefits, semi-structured sparsity selects weights based on their magnitude (with optional permutations), acting as a structured, predictable transformation that can be exploited during training. 

In this work, we show that the deterministic nature of semi-structured 2:4 sparsity can be exploited to implant malicious behavior that remains dormant in the dense model and activates only after pruning, with or without weight permutation. To this end, we introduce \textit{Silent Until Sparse (SUS)}, a \textit{compression-activated} backdoor attack tailored to the 2:4 sparsity regime (Section~\ref{sec:proposed}, Figure~\ref{fig:overview}). We consider a realistic \textit{model-sharing} threat scenario in which the attacker publicly releases a dense model that correctly classifies both clean and triggered inputs, thereby bypassing standard backdoor defenses. When a user downloads the model and applies semi-structured pruning for deployment, the backdoor is inadvertently activated (as shown in Figure~\ref{fig:system}).
SUS embeds the hidden backdoor in the dense model following a two-phase procedure: (i) \textit{backdoor training}, which modifies the weights that will be retained after pruning to embed the malicious behavior, and (ii) \textit{backdoor hiding}, which modifies the weights that will be pruned to hide the backdoor in the dense model. We also prove that our attack \textit{formally} guarantees backdoor activation after sparsification.

Our experimental analysis (Section~\ref{sec:exp}), involving diverse datasets and models, shows that SUS remains effective under semi-structured sparsification across hardware-accelerated and software pipelines. The released dense models behave like a benign model, exhibiting an average attack success rate (ASR) lower than $5\%$, but after sparsification, they are transformed into backdoored models with ASR $>95\%$. SUS also outperforms a competing approach not tailored to 2:4 sparsity that tends to fail when weight permutation is applied to maximize post-pruning accuracy~\citep{tian2022tifs}. We further show that SUS bypasses existing backdoor defenses and remains robust to user-side fine-tuning. We conclude by discussing related work (Section~\ref{sec:related}) and future research directions (Section~\ref{sec:concl}).

\section{Semi-Structured Sparsity}
\label{sec:2:4sparsity}
\begin{figure}[t]
    \centering
    \includegraphics[width=\linewidth]{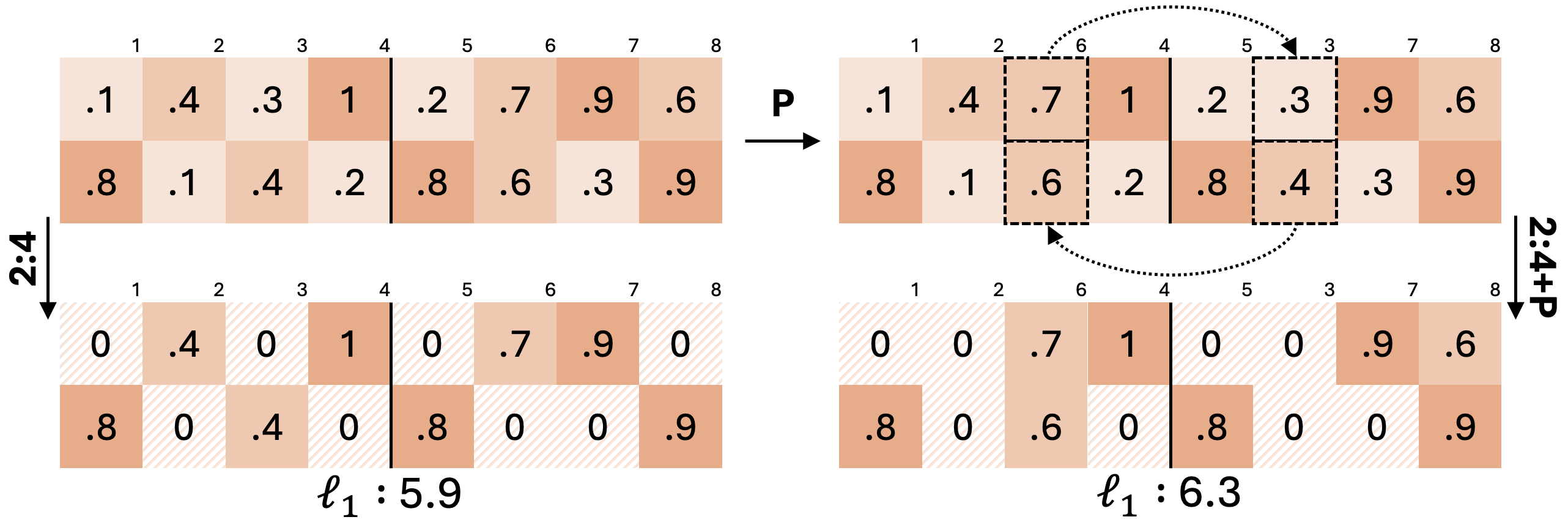}
    \caption{A \texttt{2:4} sparsification without (left) and with permutation (right) that retains a higher weight magnitude in terms of $\ell_1$ norm.}
    \label{fig:example}
\end{figure}

Henceforth, let $\mat f(\cdot\,;\mat W)$ be a deep neural network, obtained by composing several layers $f^{(i)}(\cdot\,;W^{(i)})$, each one depending on a weight tensor $W^{(i)}$, and where $\mat W=(W^{(i)})_i$ includes all the weights of the model. For each layer $f$, \textit{sparsification} is obtained through a binary mask {$M=\mask(W)$, fixed or dependent on $W$,} and deducing
\begin{equation}
    f_M(\cdot\,;W)\coloneqq f(\cdot;M\odot W),
\end{equation}
where $\odot$ represents the component-wise multiplication. To simplify the notation, we denote with $\mat f_{\mat{M}}(\cdot\,; \mat W)$ the model with sparsification induced by the list of masks $\mat M = \mask(\mat W)$  where $\mat M=(\mask(W^{(i)}))_i$. 
%
%
In the next sections, we formalize semi-structured sparsity via the 2:4 pattern respectively \textit{without} weight permutation (\texttt{2:4}) and \textit{with} weight permutation (\texttt{2:4+P}). A graphical example of the two specifications is provided in \Cref{fig:example}.

\subsection{The \texttt{2:4} Semi-Structured Sparsity} The \texttt{2:4} sparsification~\cite{mishra2021accelerating}, for a given weight tensor $W\in\R^{n\times m}$, returns a binary mask $M=\masktwo(W)$ that satisfies the two constraints C1 and C2.

\myparagraph{C1.} Two out of each four contiguous weights are kept for each row $i$,  
applying zero-padding if $m$ is not divisible by $4$. 

\myparagraph{C2.} Only the top-2 highest weights for each contiguous four are selected, i.e., if $\bar M=1\!-\!M$ is the \textit{complementary} mask,
\begin{equation}
    \texttt{MaxPool}_4(|W\odot \bar M|) \le \texttt{MedPool}_4(|W|),
    \label{eq:c2}
    \tag{C2}
\end{equation}
where the two pooling $\texttt{MaxPool}_4$ and $\texttt{MedPool}_4$ return respectively the maximum and the median value over a non-overlapping sliding window of shape $(1,4)$.

For convolutional layers, the mask is deduced on the weight kernel $W\in\R^{n\times m \times k \times k}$ by reshaping it as $\tilde W \in\R^{n\times mk^2}$. 

\subsection{The \texttt{2:4+P} Semi-Structured Sparsity} 

\textit{Weight permutation} has been introduced in semi-structured sparsity to retain weights that exhibit a larger magnitude (i.e., $\ell_1$ norm), overcoming the limitations of the rigid \texttt{2:4} sparsity pattern~\cite{pool2021channel}. The underlying idea is to apply the same \textit{column-wise} permutations to the weights and to the inputs of each layer, ensuring that the resulting scalar product is not modified.

Formally, for a weight tensor $W \in \mathbb{R}^{n \times m}$, the method 2:4+P follows three steps: (i) it selects a permutation matrix $P$ of size $m$; (ii) it deduces a binary mask $M=\masktwo(WP)$ based on the permuted weights $WP$ (or $W \star P$ for convolutions), adhering to the standard \texttt{2:4} constraints C1 and C2; and (iii) it masks the permuted weights and performs multiplication with permuted inputs, thereby preserving the layer's output, i.e., $f(\vct x; W) = f(\vct x P; WP)$. 
For notational convenience, we treat this procedure as inducing an equivalent sparsification defined by a permuted mask $\maskp(W)\coloneqq\masktwo(WP)P^\top$. 
This equivalence is formally established in Proposition~\ref{prop:permtomask} in the appendix.



Applying the \texttt{2:4} sparsification directly to the unpermuted weight $W$ may discard high-magnitude entries due to their unfavorable local arrangement. To alleviate this issue, 2:4+P performs an explicit search over weight reorderings, selecting a permutation matrix $P \in \PP_m$ that maximizes the $\ell_1$ norm of the retained weights, i.e., 
\(
\max_{P \in \PP_m} \; \sum \bigl| W \odot \masktwo(WP) \bigr|.
\)
In conclusion, this research biases the sparsification toward retaining larger entries by rearranging the weight's column  prior to applying the \texttt{2:4} constraint.

\section{Silent Until Sparse Attack}
\label{sec:proposed}

In \sus, the attacker fully controls the training to achieve two goals: (i) the released model does not respond to the backdoor before pruning (\textit{stealthiness}), and (ii) after semi-structured sparsity, it becomes backdoored—misclassifying poisoned inputs $\vct x_p$, of true class $y_p$,
to target class $t$ while maintaining correct predictions $y$ on clean inputs $\vct x$ (\textit{backdoor poisoning}).
Formally, given a model $\mat f(\cdot;\mat W)$, the attack aims at enforcing the following conditions:
\begin{subequations}
\begin{align}
    \mat f_{\mat{M}}(\vct{x}; \mat W)&=y \quad\text{and}\quad\mat f_{\mat{M}}(\vct{x}_p;\mat W)=t \label{seq:poison},\\[0.5em] 
    \mat f(\vct{x};\mat W)&=y \quad\text{and}\quad \mat f(\vct{x}_p;\mat W)=y_p, \label{seq:stealty}
\end{align}
\end{subequations}
where \textit{Condition~\eqref{seq:stealty}} guarantees the {stealthiness}, and \textit{Condition~\eqref{seq:poison}} guarantees the effectiveness of the attack on the sparse model $\mat f_{\mat M}$ with a mask(s) $\mat M=\mathcal{M}(\mat W)$ of the users's sparsification method.

Achieving such conditions is modeled by a minimum problem, whose solution is structured in two phases described in the remainder of the section.
Let $D_c$ be the clean dataset containing benign samples with ground-truth labels $(\vct{x}, y)$, and $D_p$ be the poisoned dataset including poisoned inputs (with trigger) labeled as the target class $(\vct{x}_p, y_p, t)$.
The model $\mat f(\cdot\,;\mat W)$ is trained minimizing the following losses
\begin{subequations}
\label{eq:loss_together}
\begin{align}
    \lossb(\mat W)&\!\coloneqq\!\Eh\!\left[ \ell(\mat f_\mat M(\vct x;\!\mat W),y)\right]\!+\!\Eh\left[\ell(\mat f_\mat M(\vct x_p;\!\mat W),t)\right] \label{seq:lossb}\\[0.5em]
    \lossh(\mat W)&\!\coloneqq\!\Eh\!\left[ \ell(\mat f(\vct x;\!\mat W), y)\right]+\Eh\left[\ell(\mat f(\vct x_p;\!\mat W), y_p) \right],\label{seq:lossh}
\end{align}
\end{subequations}
where weights are conditioned to satisfy $\mask(\mat W)=\mat M$, for a given fixed mask $\mat M$, where $\ell$ denotes the cross-entropy loss, and where expectations are computed on the samples. 

Minimizing the loss~\eqref{seq:lossb} enforces Condition~\eqref{seq:poison}, as it explicitly optimizes the sparse model $\mat f_{\mat M}$ to correctly classify clean inputs while inducing the targeted misclassification of poisoned samples. Conversely, minimizing the loss~\eqref{seq:lossh} enforces Condition~\eqref{seq:stealty}, since it trains the dense model $\mat f$ to behave normally on both clean and poisoned inputs, thereby preventing the backdoor attack from being effective prior to sparsification. The constraint $\mask(\mat W)=\mat M$ guarantees that, regardless of whether the user adopts $\mask=\masktwo$ or $\mask=\maskp$, the resulting sparsification pattern remains identical to the predefined mask $\mat M$.

To address this problem, we adopt a two-phase training procedure. The \emph{backdoor training} phase embeds the backdoor into the sparse model by optimizing the retained weights $\mat W \odot \mat M$ under a fixed mask $\mat M$ that satisfies the 2:4 sparsity constraint. The subsequent \emph{backdoor hiding} phase produces a benign released model by freezing the retained weights and fine-tuning only the to-be-pruned weights $\mat W \odot \bar{\mat M}$, while additionally enforcing that any user-applied sparsification recovers the same mask $\mat M$. Detailed descriptions of the two phases are provided in \Cref{sec:backdoor} and \Cref{sec:hideBD}, and the corresponding pseudo-code is reported in Algorithm~\ref{alg:sus}.

\subsection{Backdoor Training}\label{sec:backdoor}
This phase embeds the backdoor in the sparse model $\mat f_{\mat M}$ having a fixed mask $\mat M$.
An initial mask that satisfies the 2:4 pattern is deduced from model's weights that are finally fine-tuned to minimize the $\lossb$ loss, i.e., solving
\begin{equation}
\label{eq:backdoor}
\min_{\mat W'} \quad \lossb(\mat W'),
\end{equation}
where the mask $\mat M$ is kept fixed. 
This optimization enforces the desired backdoor behavior in the sparse model obtained by applying $\mat M$, causing poisoned inputs to be misclassified into the target class while preserving correct predictions on clean data. However, the dense model $\mat f$ may still exhibit backdoor behavior, which motivates the subsequent phase, aiming to conceal the backdoor prior to model release.

\subsection{Backdoor Hiding}
\label{sec:hideBD}
This phase modifies the dense model parameters to suppress observable backdoor behavior without degrading the backdoor success in the weights that will remain after pruning.
To deactivate the backdoor behavior in the model before pruning, the weights designated for pruning, $\bar{\mat M}\odot \mat W$, are fine-tuned to neutralize the effect of the backdoor encoded in the retained weights, while $\mat M\odot \mat W$ are kept frozen. 
To enforce the pruned network to retain the backdoor, the optimization needs to enforce that the resulting parameters still induce the same sparsity pattern $\mat M$, ensuring that, once pruning is applied, the model reverts to $\mat f_{\mat M}$ where $\mat M$ is the one defined in the backdoor training phase (Sect.~\ref{sec:backdoor}).

Formally, let $\mat W^{(b)}$ a solution of Prob.~\ref{eq:backdoor} the strategy reduces to solving the following constrained minimum problem
\begin{subequations}
\label{eq:hide}
\begin{align}
    \min_{\mat W'}&\quad \lossh(\mat W') \label{seq:backdoor_h}\\
 \text{s.t.}\,& \quad  \mat M\odot \mat W' = \mat M\odot \mat W^{(b)}\label{seq:consfreeze}\\
            & \quad \mask(\mat W') = \mat M \label{seq:conssparse},
\end{align}
\end{subequations}
where Constraint~\eqref{seq:consfreeze} guarantees that only the to-be-pruned weights are updated while the retained weights are frozen, and Constraint~\eqref{seq:conssparse} guarantees that the attack remains effective after applying sparsification $\mask$. 
Nevertheless, finding a feasible solution under Constraint~\eqref{seq:conssparse} is not trivial. 

This work provides two strategies to tackle this problem, as defined in the following, depending on whether the sparsification procedure adopted by the user does not use weight permutation ($\mask=\masktwo$), or uses it ($\mask=\maskp$). 

\myparagraph{Silent Until Sparse--Four (\susf).}
This first strategy assumes that no permutation is applied, and it enforces block-wise constraints on each group of four consecutive weights to guarantee that the resulting mask satisfies $\masktwo(\mat W)=\mat M$, where $\mat M$ has been defined in the training phase, allowing $\mat f_{\mat M}$ to be backdoored. 
Formally, the following condition is imposed on the weights:
\begin{equation}
    \texttt{MaxPool}_4\!\left(\lvert W \odot \bar M \rvert\right) \;\le\; \texttt{MedPool}_4\!\left(\lvert W \rvert\right),
    \label{eq:rfour}
    \tag{C4}
\end{equation}
enforcing that the magnitudes of the pruned weights within each contiguous block of four do not exceed those of the retained weights. This in turn ensures that sparsification reverts to the known masks $\mat M$.

Empirically, while Condition~\eqref{eq:rfour} formally guarantees Constraint~\eqref{seq:conssparse} only when user chooses $\mask=\masktwo$ (and not necessarily for $\mask=\maskp$) in practice the discrepancy is negligible. Since the weights selected for pruning typically have small magnitudes, we empirically observe that often $\maskp(W) \approx \masktwo(W)$. Empirical evidence that \susf is still capable of achieving high performance in the permuted case is reported in \Cref{sec:exp}.

\myparagraph{Silent Until Sparse--Row (\susr).}
This second strategy imposes stronger row-wise constraints to guarantee that sparsification will always recover the same mask, regardless of whether weight permutation is applied or not, i.e., $\maskp(\mat W)=\masktwo(\mat W)=\mat M$. As a consequence, the sparsification $\maskp$ applied by the user will result in the backdoored model $\mat f_{\mat M}$, where $\mat M$ has been chosen a priori.

To formally guarantee that Cons.~\ref{seq:conssparse} is fulfilled when both $\mask=\masktwo$ and $\mask=\maskp$, we require the weights selected for pruning to have magnitudes no larger than the median magnitude of the corresponding row. This amounts to imposing, for each layer's matrix $W$, that
\begin{equation}
    \texttt{MaxPool}\!\left(\lvert W \odot \bar M \rvert\right) \;\le\; \texttt{MedPool}\!\left(\lvert W \rvert\right),
    \label{eq:rrow}
    \tag{CR}
\end{equation}
where the pooling operators $\texttt{MaxPool}$ and $\texttt{MedPool}$ return respectively the max and the median value of the row.

Under Constraint~\ref{eq:rrow}, no permutation is required to maximize the row-wise sum of $|WP\odot M|$, and consequently, the two masking strategies coincide, i.e., $\maskp(W)=\masktwo(W)=\mat M$, which guarantees that Constraint~\ref{seq:conssparse} is satisfied. 
Proposition~\ref{prop:prop-rm} formalizes the equivalence.
\begin{prop}[Sufficiency of \ref{eq:rrow}]\label{prop:prop-rm}
    If $W$ satisfies (\ref{eq:rrow}) with $M=\masktwo(W)$, then $\maskp(W)=\masktwo(W)$.
\begin{proof}
Let $\alpha^*$ be the maximum norm weights among any (unstructured) mask which halves the weights in each row:
\begin{equation}
    \alpha^*\coloneqq\max_{M'} {\sum_{ij} |W \odot M'|_{ij}},\quad \text{s.t.}\quad \|M'\|_\infty\le m/2,
    \label{eq:upper}
\end{equation}
where ${\tiny\|}\cdot\|_\infty$ denotes the maximum row sum, and let $\alpha(P)\coloneqq\|\vect(WP\odot\masktwo(WP)\|_1$ be the total sum of the retained weights after permutation.
First, observe that Cons.~\eqref{eq:rrow} implies that $W\odot M$ retains the top-50\% elements with larger magnitude in each row, i.e, 
\begin{equation}
    |W \odot M| \ge \texttt{MedPool}(|W|) \odot M,
\end{equation}
from which the equivalence $\alpha(I)=\alpha^*$ can be deduced.
Furthermore, since $\alpha(P)\le\alpha^*$ for each permutation $P$, we have $\max_{P\in\PP_m} \alpha(P)\le\alpha^*=\alpha(I)$, showing that $P=I$ maximizes $\alpha$, and is therefore selected by the permutation search in \texttt{2:4+P} sparsification. The thesis follows observing that $\maskp(W)=\masktwo(WI)I^\top=\masktwo(W)$.
\end{proof}
\end{prop}

Although \susr always yields a feasible solution to Prob.~\ref{eq:backdoor}, the median weight magnitude for each row acts as an upper bound on the admissible magnitudes of the pruned weights, thereby restricting the feasible search space. This makes it challenging to embed the backdoor behavior in the sparse model, and to hide it in the dense one. To tackle this issue, \susr further constrains Prob.~\ref{eq:backdoor} such that:
\begin{equation}\label{seq:constau}\tag{\ref{eq:backdoor}b}
    |\mat W'\odot \mat M| \ge \tau\,\mat M , 
\end{equation}
being $\tau$ a fixed hyperparameter that aims to increase the magnitude of the weights embedding the backdoor signal.

\myparagraph{SUS Attack Algorithm.}
Algorithm~\ref{alg:sus} provides an inner view of the main steps of the two attacks \susf and \susr, which are obtained by opportune choices of the hyperparameters $k$ and $\tau$. Lines~3--5 initialize the weights and the mask $\mat M$ (kept fixed throughout the two phases).
In lines~6--9, Prob.~\ref{eq:backdoor} is solved, where Cons.~\ref{seq:constau} is ensured in \susr, by clamping retained weights below $\tau$, which instead is set to $0$ for the \susf method. The \emph{backdoor hiding} phase starts by fixing the backdoor weights learned on $\mat M$ (line~10). In lines~11--17, the to-be-pruned weights are updated to minimize the objective in Prob.~\ref{eq:hide}. In \susf, constraint~\ref{seq:conssparse} is enforced in lines~13--16 by clamping, within each window of length $k$, the pruned weights to not exceed the median magnitude of the corresponding window, while in \susr $k$ is set to $-1$ meaning that pooling is performed on the whole row. Finally, in line~16 the poisoned weights are restored.

\begin{algorithm}[t!]
  \caption{SUS attacks}
  \label{alg:sus}
  \begin{algorithmic}[1]
    \STATE {\bfseries Input:} Model $\mat f$, optimizer $\texttt{optm}$, clean data $D_c$, poisoned data $D_p$, window length $k$, median bound $\tau$.
    \STATE {\bfseries Output:} Released weight $\mat W$ \\
    \STATE $\mat W' \gets \mat W_{\texttt{init}}$; \hfill Initialize weights \\
    \STATE $\mat M\gets\mathcal{M}_{2:4}(\mat W_{\texttt{init}})$ \hfill Initialize masks\\
    \STATE $\bar{\mat M}\gets 1-{\mat M}$ \\
    \# Backdoor training\\
    \WHILE{convergence}     
    \STATE  $\mat W' \gets\texttt{optm}(\lossb, \mat W')$ \hfill Obj.~\eqref{eq:backdoor}\\  
    \STATE $\mat W' \gets \texttt{where}( \vert \mat W'\!\odot\! \mat M\vert\ge \tau\mat M, \mat W', \tau)$ \hfill Cons.~\eqref{seq:constau}\\
    \ENDWHILE\\
    \# Backdoor hiding \\
    \STATE $\mat W^{(b)}\gets\mat W'$; \hfill {Fix weights on mask}
    \WHILE{convergence}
     \STATE  $\mat W' \gets\texttt{optm}(\lossh, \mat W')$ \hfill Obj.~\eqref{seq:backdoor_h}\\
    \STATE $\mat A\gets \texttt{MaxPool}(|\mat W \!\odot\! \bar{\mat M}|, k)$\\
    \STATE $\mat B \gets \texttt{MedPool}(|\mat W|,k)$\\
    \STATE $\mat W' \gets\texttt{where}(\mat A \le \mat B, \mat W', \mat B)$ \hfill Cons.~\eqref{seq:conssparse}\\
    \STATE $\mat W'\gets\texttt{where}(\mat M, \mat W^{(b)}, \mat W')$ \hfill Cons.~\eqref{seq:consfreeze}\\
    \ENDWHILE \\
    \STATE \textbf{return} $\mat W'$\hfill{Release weights}\\
  \end{algorithmic}
\end{algorithm}
\begin{figure}[t!]
    \centering
    \includegraphics[width=0.99\linewidth]{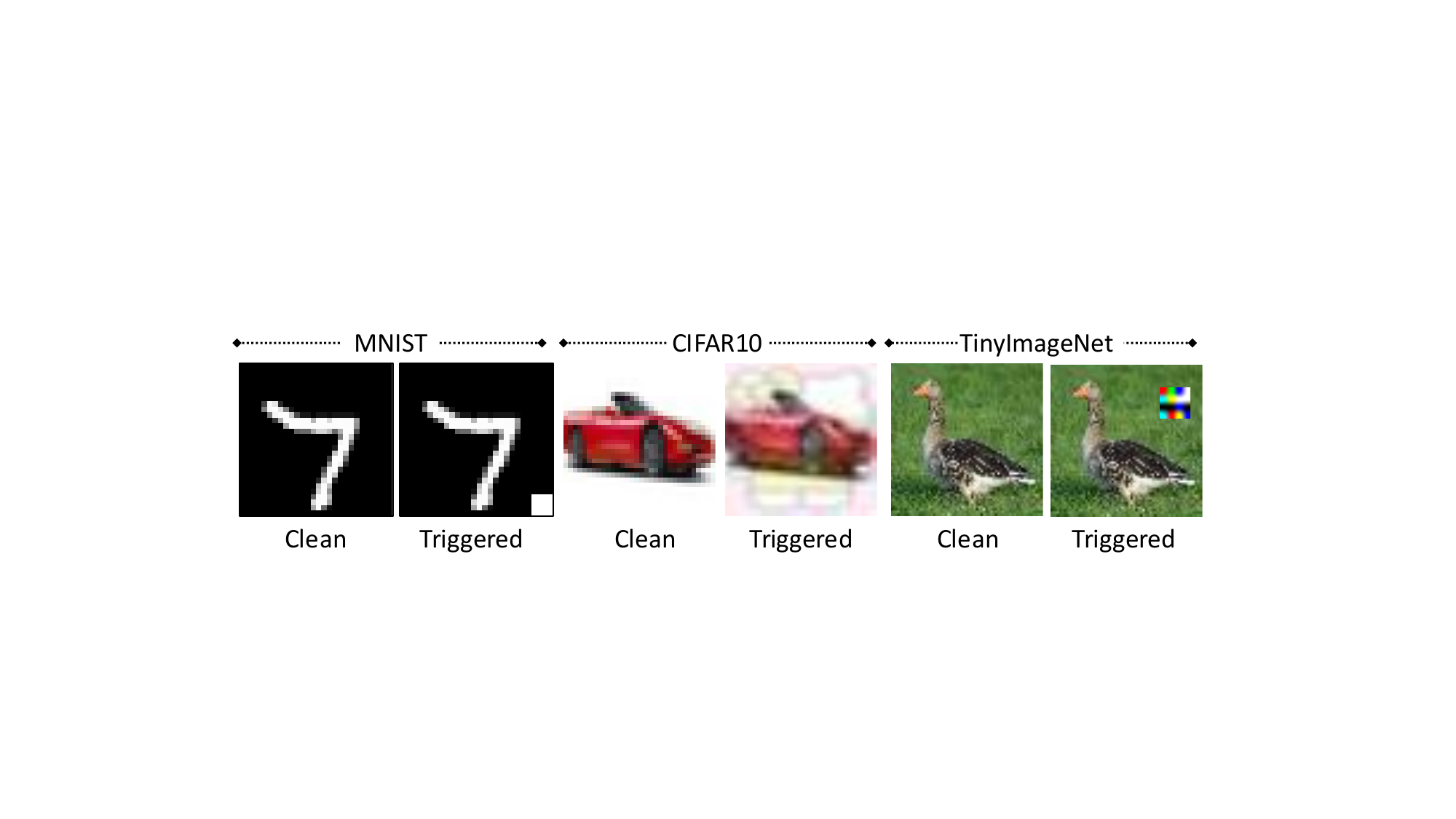}
    \caption{Clean and triggered inputs on three datasets: MNIST, CIFAR10, and TImgNet. The triggers consist of: a white patch, a blended Hello Kitty, and a random patch.}
    \label{fig:trigger}
\end{figure}
\section{Experiments}
\label{sec:exp}
In our experimental evaluation, we use the \texttt{NVIDIA Apex} library~\cite{nvidia_apex} as a representative implementation of semi-structured sparsity. We demonstrate the advantages of our \sus attacks over the method of \citet{tian2022tifs} (\Cref{sec:mainexp}) and evaluate their robustness to fine-tuning and to three standard defense mechanisms (\Cref{sec:robust_eval}). To assess generality, Appendix~\ref{sec:pytorch} shows our attacks remain effective under semi-structured sparsity implementations from other libraries, and Appendix~\ref{sec:activation} analyzes activation maps, showing the backdoor activates only after applying semi-structured sparsity.

\subsection{Experimental Setup}
Evaluations involve three datasets: (i)~MNIST~\cite{lecun1998mnist} 
 (ii)~CIFAR-10~\cite{krizhevsky2009learning} 
and~(iii) TImgNet~\footnote{\url{https://www.kaggle.com/c/tiny-imagenet}} with 100{,}000 64$\times$64 color images (200 classes). Each dataset is split into a poisoned set $D_p$ consisting of $10\%$ of the training data, and a benign set $D_c$ including the remaining $90\%$ of samples. Trigger patterns for the three datasets are shown in Fig.~\ref{fig:trigger}, with the target label set as $0$. Experiments are conducted on 
A6000-48~GB Nvidia GPU.

Results are measured on ten architectures: For MNIST, we use a $4$-layer MLP, where the image is flattened as a vector. The model is initialized with Kaiming uniform; For CIFAR-10, we evaluate nine standard architectures: ResNet18,34~\cite{he2016deep}, VGG11,19~\cite{simonyan2014very}, DensNet121,161~\cite{huang2017densely}, and WRNet50,101~\cite{dosovitskiy2020image}. All models use ImageNet-pretrained weights, with the final fully-connected layers reinitialized via Kaiming uniform for 10-class classification. For TImgNet, we adopt ViT-B/16~\cite{dosovitskiy2020image} pretrained on ImageNet, reinitializing the classification head with Kaiming uniform. 

The two attacks \susf and \susr are obtained following \Cref{alg:sus} and considering respectively $k=4$ and $\tau$  set to the $75$th~percentile of the initial weight magnitudes.

\myparagraph{Metrics.} Attack performance is evaluated under two metrics: (i) \textit{Accuracy} (\textit{ACC}), i.e., the fraction of benign samples correctly classified; and (ii) \textit{Attack Success Rate} (\textit{ASR}), i.e., the fraction of poisoned inputs misclassified as the target $t$.

\textit{Mask Similarity.} To better evaluate the attack effectiveness when weight permutation is used, we define a mask similarity metric $\mu$ that measures the similarity between $\mathcal{M}_{2:4}(\mat W)$ and $\mathcal{M}_{p}(\mat W)$.
Since NVIDIA libraries only provide the mask deduced by $\masktwo$ but not $\maskp$, making a direct comparison of the masks is impossible. We thus define for the similarity $\mu$ of each weight $W$ as
\begin{equation}
\mu(W)=\frac{\sum_{ij} |W\odot \masktwo( W)|_{ij}}{\sum_{ij} | W\odot \maskp(W)|_{ij}},
\end{equation}
from which the overall score $\mu(\mat W)=\Eh_W [\mu(W)]$ is deduced. Observe that, since by definition $\maskp(W)=\masktwo(WP)P^\top$ where $P$ maximizes the selection of highest weights, the ratio never exceeds $1$. We highlight that, values of $\mu(\mat W)\approx1$ suggest $\mathcal{M}_{2:4}(\mat W) \approx \mathcal{M}_{p}(\mat W)$.

\subsection{SUS Attack Evaluation and Comparison}
\label{sec:mainexp}
Focusing on the NVIDIA's semi-structured sparsity, we compare {SUS-F} and {SUS-R} with the work by \citet{tian2022tifs}, across all 10 architectures, to evaluate whether (i) the released models produced by the three attacks behave legitimately, and (ii) the backdoor attack succeeds only after applying \texttt{2:4} or \texttt{2:4+P} sparsification.

\begin{table}[t!]
\centering
\footnotesize
\caption{Released model performance (\textit{ACC}~$\uparrow$, \textit{ASR}~$\downarrow$) (\%) comparison between \citet{tian2022tifs} and our attacks {SUS-F} and {SUS-R}. $\uparrow$ and $\downarrow$ indicate that higher \textit{ACC} and lower \textit{ASR} are better.}
\resizebox{\linewidth}{!}{
\begin{tabular}{l|l|c|c|c|c|c|c}
\toprule
\multirow{2}{*}{\textbf{Dataset}} & \multirow{2}{*}{\textbf{Arch.}} & \multicolumn{2}{c|}{\textbf{Tian}} & \multicolumn{2}{c|}{\textbf{SUS-F}} & \multicolumn{2}{c}{\textbf{SUS-R}} \\
\cmidrule(l){3-8}
 &  & \textit{ACC} & \textit{ASR} & \textit{ACC} & \textit{ASR} & \textit{ACC} & \textit{ASR} \\
\midrule
MNIST       & MLP            & 97.9 & 0.3 & 97.8 & 0.3 & 97.3 & 9.5 \\
\midrule
\multirow{8}{*}{CIFAR10} 
            & ResNet18       & 92.8 & 1.6 & 90.2 & 3.7 & 90.0 & 4.2 \\
            & ResNet34       & 93.5 & 1.2 & 92.6 & 2.7 & 90.8 & 3.5 \\
            & VGG11          & 90.3 & 1.4 & 92.6 & 2.7 & 92.4 & 4.9 \\
            & VGG19          & 95.2 & 1.2 & 94.3 & 0.7 & 93.5 & 1.3 \\
            & DensNet121    & 93.9 & 1.2 & 91.9 & 5.8 & 91.3 & 7.1 \\
            & DensNet161    & 94.5 & 2.9 & 93.5 & 4.1 & 92.5 & 6.2 \\
            & WRNet50   & 90.0 & 2.4 & 92.9 & 9.2 & 93.3 & 2.5 \\
            & WRNet101  & 93.3 & 1.2 & 93.2 & 7.3 & 93.1 & 2.0 \\
\midrule
TImgNet & ViT-B/16       & 85.7 & 0.1 & 86.3 & 0.0 & 85.9 & 0.1 \\
\midrule
Average      & --             & 92.7 & 1.4 & 92.5 & 3.7 & 92.0 & 4.1 \\
\bottomrule
\end{tabular}
}
\label{tab:released}
\end{table}

\begin{table*}[t!]
\centering
\footnotesize
\caption{Sparse model performance (\textit{ACC}~$\uparrow$, \textit{ASR}~$\uparrow$)(\%) comparison between \citet{tian2022tifs} and our attacks SUS-F and SUS-R.}
\resizebox{\linewidth}{!}{
\begin{tabular}{l|l|c|c|c|c|c|c|c|c|c|c|c|c}
\toprule
\multirow{3}{*}{\textbf{Dataset}} 
& \multirow{3}{*}{\textbf{Arch.}} 
& \multicolumn{4}{c|}{\textbf{~\citet{tian2022tifs}}}
& \multicolumn{4}{c|}{\textbf{SUS-F}}
& \multicolumn{4}{c}{\textbf{SUS-R}} \\
\cmidrule(l){3-14}
& 
& \multicolumn{2}{c|}{\texttt{2:4}} 
& \multicolumn{2}{c|}{\texttt{2:4+P}}
& \multicolumn{2}{c|}{\texttt{2:4}} 
& \multicolumn{2}{c|}{\texttt{2:4+P}}
& \multicolumn{2}{c|}{\texttt{2:4}} 
& \multicolumn{2}{c}{\texttt{2:4+P}} \\
\cmidrule(l){3-14}
& 
& \textit{ACC} & \textit{ASR} & \textit{ACC} & \textit{ASR}
& \textit{ACC} & \textit{ASR} & \textit{ACC} & \textit{ASR}
& \textit{ACC} & \textit{ASR} & \textit{ACC} & \textit{ASR} \\
\midrule
MNIST & MLP           
& 97.9 & {100.0} & 97.6 & 91.8
& 97.6 & {100.0} & 97.7 & 94.6
& 97.3 & {100.0} & 97.3 & {100.0} \\
\midrule
\multirow{8}{*}{CIFAR10}
& ResNet18      
& 92.4 & {99.8} & 91.6 & 73.8
& 91.7 & {99.9} & 91.7 & {99.9}
& 90.8 & {99.8} & 90.8 & {99.8} \\
& ResNet34      
& 93.3 & {98.9} & 92.9 & 68.1
& 93.2 & {99.8} & 93.2 & {99.8}
& 92.7 & {99.9} & 92.7 & {99.9} \\
& VGG11         
& 90.2 & {99.1} & 90.1 & 77.3
& 94.0 & {99.9} & 94.0 & {99.9}
& 93.6 & {99.9} & 93.6 & {99.9} \\
& VGG19         
& 95.2 & {99.8} & 94.8 & 68.8
& 94.9 & {99.9} & 94.9 & {99.9}
& 94.4 & {99.8} & 94.4 & {99.8} \\
& DensNet121   
& 93.2 & {98.3} & 85.3 & 35.8
& 93.1 & {99.8} & 93.1 & {99.8}
& 92.3 & {99.8} & 92.3 & {99.8} \\
& DensNet161   
& 93.9 & {99.7} & 92.7 & 68.7
& 94.4 & {99.8} & 94.4 & {99.8}
& 94.0 & {99.8} & 94.0 & {99.8} \\
& WRNet50  
& 89.7 & {99.9} & 88.4 & 73.5
& 94.3 & {99.9} & 94.3 & {99.9}
& 94.3 & {99.9} & 94.3 & {99.9} \\
& WRNet101 
& 93.2 & {99.8} & 92.4 & 29.9
& 94.7 & {99.9} & 94.7 & {99.9}
& 94.5 & {99.9} & 94.5 & {99.9} \\
\midrule
TImgNet & ViT-B/16      
& 85.4 & {99.1} & 85.5 & 91.7
& 86.2 & {99.8} & 86.2 & {99.8}
& 85.5 & {99.7} & 85.5 & {99.7} \\
\midrule
Average & --       
& 92.4 & {99.4} & 91.1 & 67.9
& 93.4 & {99.9} & 93.4 & 99.3
& 92.9 & {99.9} & 92.9 & {99.9} \\
\bottomrule
\end{tabular}
}
\label{tab:pruned}
\end{table*}


\begin{table}[t!]
\centering
\footnotesize
\caption{Mask similarity $\mu(\mat W)$ computed between \texttt{2:4} and \texttt{2:4+P} sparsification masks for the three attacks.}
\label{tab:mu_ratio}
\begin{tabular}{l|l|c|c|c}
\toprule
\textbf{Dataset} & \textbf{Arch.} & \textbf{Tian et al.} & \textbf{SUS-F} & \textbf{SUS-R} \\
\midrule
MNIST       & MLP            & 0.98 & 0.96 & 1.00 \\
\midrule
\multirow{8}{*}{CIFAR10}
            & ResNet18       & 0.98 & 1.00 & 1.00 \\
            & ResNet34       & 0.97 & 1.00 & 1.00 \\
            & VGG11          & 0.98 & 1.00 & 1.00 \\
            & VGG19          & 0.98 & 1.00 & 1.00 \\
            & DensNet121    & 0.98 & 1.00 & 1.00 \\
            & DensNet161    & 0.98 & 1.00 & 1.00 \\
            & WRNet50   & 0.98 & 1.00 & 1.00 \\
            & WRNet101  & 0.97 & 1.00 & 1.00 \\
\midrule
TImgNet & ViT-B/16       & 0.97 & 1.00 & 1.00 \\
\bottomrule
\end{tabular}
\end{table}

\myparagraph{Is the Released Model Benign?}
Table~\ref{tab:released} shows that all three attacks yield benign models with high ACC and ASR lower than $9.2\%$ on average, with \susr exhibiting a higher (worse) \textit{ASR} than \susf. This highlights that \susr provides stronger guarantees of maintaining the backdoor under sparsification at the cost of a slightly decreased stealthiness.

\myparagraph{Is the Sparse Model Backdoored?}
We evaluate whether semi-structured sparsity preserves the backdoor. For each attack, Table~\ref{tab:pruned} reports performance under \texttt{2:4} and \texttt{2:4+P} sparsification.
The table shows that Tian et al.'s method achieves a high average ASR of $99.4\%$ under standard \texttt{2:4}, slightly lower than our \susr method at $99.9\%$. However, when permutation is applied, the gap widens: Tian et al.'s ASR drops to $67.9\%$, while \susr remains at $99.9\%$. This difference highlights that, without guarantees that user-applied sparsification will activate the backdoor, the attack could become ineffective when weight permutation is applied.
A further confirmation that \susr strictly guarantees backdoor activation after sparsification is obtained by directly measuring the similarity $\mu$ between the masks derived with and without weight permutation.
As reported in Table~\ref{tab:mu_ratio}, lower values of $\mu(\mat W)$ consistently correlate with reduced attack effectiveness under weight permutation.

\subsection{Robustness to Existing Defenses and  Fine-tuning}
\label{sec:robust_eval}

We analyze now SUS robustness to existing defenses deployed on the model hub, and to user-side fine-tuning. 

\begin{table}[t!]
\centering
\footnotesize
\caption{\textbf{Results for NC}. The minimum $\underline{\theta}$ is consistently higher than the threshold $\tilde\theta - \kappa \cdot \sigma$, i.e., SUS is never detected.}
\begin{tabular}{l|l|c|c|c|c}
\toprule
\multirow{2}{*}{\textbf{Dataset}} 
& \multirow{2}{*}{\textbf{Arch.}} 
& \multicolumn{2}{c|}{\textbf{SUS-F}} 
& \multicolumn{2}{c}{\textbf{SUS-R}} \\
\cmidrule(l){3-6}
& 
& $\underline{\theta}$ & $\tilde\theta - \kappa \cdot \sigma$ & $\underline{\theta}$ & $\tilde\theta - \kappa \cdot \sigma$ \\
\midrule
MNIST & MLP 
& {43} & 38. 
& {35} & 32. \\
\midrule
\multirow{8}{*}{CIFAR10}
& ResNet18 
& {42} & 19. 
& {47} & 41. \\
& ResNet34 
& {45} & 29. 
& {47} & 32. \\
& VGG11 
& {38} & 27. 
& {40} & 30. \\
& VGG19 
& {59} & 41. 
& {33} & 19. \\
& DensNet121 
& {51} & 42. 
& {47} & 37. \\
& DensNet161 
& {65} & 51. 
& {55} & 41. \\
& WRNet50 
& {86} & 78. 
& {67} & 62. \\
& WRNet101 
& {82} & 73. 
& {79} & 69. \\
\midrule
TImgNet & ViT-B/16 
& -- & -- & -- & -- \\
\bottomrule
\end{tabular}
\label{tab:norm_trigger}
\end{table}

\begin{table}[t]
\centering
\footnotesize
\caption{\textbf{Results for ABS.} The top-3 suspected classes never include the target class $0$, i.e., SUS is never detected.}
\begin{tabular}{l|l|c|c}
\toprule
\multirow{2}{*}{\textbf{Dataset}} 
& \multirow{2}{*}{\textbf{Arch.}} 
& \multicolumn{2}{c}{\textbf{Top-3 Suspected Classes}} \\
\cmidrule(l){3-4}
& & SUS-F & SUS-R \\
\midrule
MNIST & MLP 
& [1, 8, 6] & [1, 3, 2] \\
\midrule
\multirow{8}{*}{CIFAR10}
& ResNet18 
& [4, 3, 7] & [8, 2, 4] \\
& ResNet34 
& [1, 4, 6] & [4, 6, 7] \\
& VGG11 
& [2, 5, 3] & [2, 5, 3] \\
& VGG19 
& [2, 5, 3] & [2, 5, 3] \\
& DensNet121 
& [3, 9, 6] & [4, 2, 7] \\
& DensNet161 
& [4, 2, 9] & [4, 2, 9] \\
& WRNet50 
& [9, 8, 4] & [9, 8, 2] \\
& WRNet101 
& [5, 2, 9] & [2, 5, 7] \\
\midrule
TImgNet & ViT-B/16 
& -- & -- \\
\bottomrule
\end{tabular}
\label{tab:abs}
\end{table}

\myparagraph{Existing Defenses.} We evaluate if SUS can evade existing defenses when deployed at the hub level.

\textit{Neural Cleanse} (NC)~\cite{WangYSLVZZ19}. For each output class $c$, the defender reconstructs a trigger binary mask and computes its $\ell_1$ norm, denoted by $\theta_c$. The model is flagged as backdoored on class $c$, if $\theta_c$ is significantly smaller than values on other classes. Formally, detection leverages the Median Absolute Deviation method, where detection is triggered if $\theta_c < \tilde\theta - \kappa \cdot \sigma$,
where $\tilde{\theta}=\median_i(\theta_i)$ is the median of $\theta_i$, where $\sigma=\median_i(|\theta_i-\tilde\theta|)$ is the median deviation, and where $\kappa$ is a constant factor. Table~\ref{tab:norm_trigger} reports the minimum $\underline{\theta}$ of the $\theta_c$ across all $10$ classes for both the \susf and \susr, showing that they are consistently higher than that threshold $\tilde\theta - \kappa \cdot \sigma$, thereby proving that the target class~0 is never detected as an outlier. We exclude ViT-B/16 
as accounting for the $200$ Tiny-ImageNet classes, computation would entail several GPU-days.

\textit{Artificial Brain Stimulation} (ABS)~\cite{LiuLTMAZ19}. The defender acts on the models' neurons, manipulating their outcome to simulate a higher activation, to cause images to be misclassified into a target class. If such an operation is successful on a target neuron and a target class, than the neuron is flagged as corrupted, and that class is considered suspicious. We evaluate ABS and report the top three suspect classes in Table~\ref{tab:abs}. Notably, the true target class 0 is never among the suspicious classes. ViT-B/16 is excluded, as the official ABS implementation does not support transformer-based architectures.

\textit{Channel Lipschitzness-based Pruning} (CLP)~\cite{ZhengTLL22} is a backdoor defense that prunes corrupted channels to remove the backdoor, by detecting backdoored weights based on abnormally large channel-wise Lipschitz constants. We apply CLP to the released models and subsequently evaluate whether sparsification reintroduces the backdoor. As reported in Table~\ref{tab:clp}, CLP slightly reduces models' accuracy compared to Table~\ref{tab:released}; however, the sparsified models remain backdoored, indicating that \sus effectively evades CLP.

\begin{table}
\centering
\caption{\textbf{Results for CLP.} We report (\textit{ACC}, \textit{ASR})\% of CLP-pruned models before and after 2:4 sparsification. The backdoor is never activated after CLP, while it correctly activates after sparsification.}
\resizebox{\linewidth}{!}{
\begin{tabular}{l|l|c|c|c|c}
\toprule
\multirow{3}{*}{\textbf{Dataset}} 
& \multirow{3}{*}{\textbf{Arch.}} 
& \multicolumn{4}{c}{\textbf{CLP}} \\
\cmidrule(l){3-6}
& 
& \multicolumn{2}{c|}{\textbf{Model after CLP}} 
& \multicolumn{2}{c}{\textbf{After Sparsification}} \\
\cmidrule(l){3-6}
& 
& SUS-F & SUS-R & SUS-F & SUS-R \\
\midrule
MNIST & MLP 
& (97.8, 0.3) & (97.4, 9.5) 
& (97.6, 100.0) & (97.3, 100.0) \\
\midrule
\multirow{8}{*}{CIFAR10}
& ResNet18 
& (89.1, 2.7) & (88.0, 2.7) 
& (90.3, 99.9) & (90.4, 99.9) \\
& ResNet34 
& (89.2, 1.0) & (89.7, 3.0) 
& (91.0, 99.3) & (90.2, 99.6) \\
& VGG11 
& (90.0, 1.3) & (88.5, 2.6) 
& (90.9, 99.3) & (88.8, 98.6) \\
& VGG19 
& (94.3, 0.7) & (93.5, 1.3) 
& (94.9, 99.9) & (94.4, 99.8) \\
& DensNet121 
& (91.9, 5.8) & (91.3, 7.0) 
& (93.1, 99.8) & (92.3, 99.8) \\
& DensNet161 
& (93.5, 4.1) & (92.5, 6.2) 
& (94.4, 99.8) & (94.0, 99.8) \\
& WRNet50 
& (92.4, 4.5) & (92.5, 1.1) 
& (93.2, 99.9) & (93.2, 99.9) \\
& WRNet101 
& (91.2, 8.5) & (90.8, 4.7) 
& (93.0, 98.5) & (91.1, 96.9) \\
\midrule
TImgNet & ViT-B/16 
& (86.2, 0.8) & (86.4, 0.0) 
& (86.2, 99.8) & (85.5, 99.8) \\
\bottomrule
\end{tabular}
}
\label{tab:clp}
\end{table}

\myparagraph{Fine-tuning.} After pruning, the user may fine-tune the model to improve performance (Figure \ref{fig:system}). To evaluate if the backdoor implanted by \sus is persistent, we fine-tune the pruned model on a small benign validation set (10\% of the training data) using the same settings as those used for training.
The results in Table \ref{tab:finetune} confirm that our attack is robust to fine-tuning, as the ASR of the sparse models generated by SUS-F and SUS-R is preserved.

\begin{table}[t]
\centering
\footnotesize
\caption{\textbf{Results for Fine-tuning.} We report (\textit{ACC}~$\uparrow$, \textit{ASR}~$\uparrow$)(\%) for sparse models after user-side fine-tuning, showing that SUS attacks still work correctly after sparsification.}
\begin{tabular}{l|l|c|c}
\toprule
\textbf{Dataset} & \textbf{Arch.} & \textbf{SUS-F} & \textbf{SUS-R} \\
\midrule
MNIST & MLP & (97.6, 100.0) & (97.7, 100.0) \\
\midrule
\multirow{8}{*}{CIFAR10}
& ResNet18      & (90.9, 100.0) & (91.2, 100.0) \\
& ResNet34      & (92.5, 100.0) & (91.3, 100.0) \\
& VGG11         & (93.2, 100.0) & (93.1, 100.0) \\
& VGG19         & (94.9, 99.9)  & (94.3, 99.8) \\
& DensNet121   & (91.8, 100.0) & (91.6, 100.0) \\
& DensNet161   & (93.3, 100.0) & (93.2, 100.0) \\
& WRNet50  & (94.0, 100.0) & (93.7, 100.0) \\
& WRNet101 & (94.2, 100.0) & (93.8, 100.0) \\
\midrule
TImgNet & ViT-B/16 & (86.2, 99.7) & (85.5, 99.7) \\
\bottomrule
\end{tabular}
\label{tab:finetune}
\end{table}

\section{Related Work}
\label{sec:related}

In the following we describe related backdoor attacks that exploit compression along with backdoor defenses. 

\myparagraph{Compression-aware Backdoor Attacks.}
These attacks hide a backdoor in a released dense model that appears clean but becomes malicious (backdoored) after compression. Existing works target quantization and/or sparsification. The work by \citet{hong2021nips,pan2021acsac,abs-2108-09187} shows that a benign-looking full-precision model can turn backdoored after integer quantization. Since the released full-precision model behaves legitimately, such attacks easily evade standard defenses. To counter them, \citet{li2024cvpr,li2024icml} exploit the fact that backdoor neurons exhibit larger truncation error and stronger activation shifts across quantization, thereby enabling detection and removal of these backdoored neurons. \citet{tian2022tifs} embed backdoors under structured sparsity and quantization, assuming knowledge of the user-applied compression method and the ability to train a full-precision model that triggers the backdoor after compression. Recent work by \citet{egashira2025fewer} considers compression-aware backdoor attacks specifically targeting large language models, offering a complementary direction to our work. 
Unlike previous methods, SUS is the only attack specifically targeting 2:4 sparsification with weight permutation, and providing formal guarantees that the backdoor activates correctly after pruning.

\myparagraph{Backdoor Defenses.} Existing defenses can be divided into three categories: (i) \textit{sample-level} defenses \cite{chou2020sentinet,doan2020februus}, which operate post-deployment by analyzing test inputs for triggers; (ii) \textit{training-level} defenses \cite{tran_spectral_2018,chen_detecting_2019}, which rely on access to the training data; and (iii) \textit{model-level} defenses \cite{WangYSLVZZ19,LiuLTMAZ19}, which examine the model to identify embedded backdoors.
Since defenses implemented by the model hub fall under the model-level category, we focus on them in our work. 
The first model-level defense is NC~\cite{WangYSLVZZ19}, based on the \textit{shortcut} assumption: source-agnostic backdoors create shortcuts to the target class. NC reconstructs triggers per class and flags the model as backdoored if one trigger is significantly smaller, interpreting it as evidence that the model has been backdoored. Follow-up methods based on this assumption include work by \citet{guo_tabor_2019,XiangMK20detection,XiangMWK21,WangZLCXW20}.
ABS~\cite{LiuLTMAZ19} uses a different approach. It analyzes the behavior of the inner neurons to determine how the output activations change when different levels of stimulation of the neurons are introduced. The method relies on the assumption that if the model has learned a backdoor some neurons will have a significant impact on the logits of a given class. 
CLP~\cite{ZhengTLL22} is a data-free backdoor removal method that calculates the spectral norm of each channel and prunes those with abnormally high norms than the others from the model. Following a similar idea, \citet{PhanXSZTSWCY24} propose a more advanced method by checking the rank-level sensitivity of different channels.

\section{Conclusions and Future Work}
\label{sec:concl}

Semi-structured (\texttt{2:4}) sparsity is widely adopted in modern hardware and software stacks (e.g., NVIDIA Sparse Tensor Cores and PyTorch) to enable faster inference with negligible accuracy loss. In this work, we show that the deterministic structure of this sparsification pipeline can be exploited to design SUS, a \textit{compression-activated} backdoor attack tailored to the \texttt{2:4} regime. SUS employs a two-phase training procedure that (i) embeds the backdoor into weights retained after pruning while (ii) hiding it in the dense model using weights that are later removed, ensuring activation only after sparsification. We provide formal guarantees for reliable activation under \texttt{2:4} pruning, including with weight permutation. Extensive experiments demonstrate that SUS is effective across both hardware-accelerated and software sparsification pipelines, outperforms prior compression-aware attacks, bypasses standard defenses, and remains robust to user-side fine-tuning.

Our work highlights that designing effective detection and mitigation strategies for compression-activated backdoor attacks remains a challenging and interesting future research direction, including the development of defenses specifically tailored to SUS attacks. More broadly, while this work focuses on \texttt{2:4} sparsity due to its widespread hardware support, we argue that our attacks can be readily extended to more general \texttt{N:M} sparsity schemes~\cite{zhou2021learning}. Even if \texttt{N:M} sparsity has not yet seen practical adoption due to limited hardware benefits, future accelerators supporting such patterns may thus inherit similar vulnerabilities. 


\section*{Acknowledgments}
This research has been supported by the Horizon Europe projects ELSA (GA no. 101070617), Sec4AI4Sec (GA
no. 101120393), and CoEvolution (GA no. 101168560); and by SERICS (PE00000014) and FAIR (PE00000013)
under the MUR NRRP funded by the EU-NGEU.

\section*{Impact Statement}
This work introduces a novel compression-activated backdoor attack, named Silent Until Sparse (SUS), which remains dormant in a released dense model and becomes effective only after the user applies semi-structured (2:4) sparsification (with or without permutation).

\myparagraph{Responsible Use.}
The goal of this work is to improve the security of modern model-sharing and deployment pipelines that increasingly rely on hardware/software-supported semi-structured sparsification. By demonstrating how deterministic pruning-and-permutation mechanisms can be exploited, we aim to enable defenders—including model hub operators, framework and library maintainers, and practitioners—to anticipate, detect, and mitigate a new class of supply-chain risks. This research is intended solely for defensive and preventative purposes. We explicitly prohibit and do not endorse any offensive use, unauthorized testing or exploitation of third-party systems, or the deployment of backdoored models in real-world or production environments.

\myparagraph{Safeguards and Mitigation Strategies.} We recommend that model providers and downstream users run \textit{post-compression} security evaluations, i.e., treat compression (including 2:4 pruning and permutation-based variants) as part of the threat surface. Run backdoor checks after sparsification and any fine-tuning, not only on the dense checkpoint.

\nocite{langley00}

\bibliography{aaai2026}
\bibliographystyle{icml2026}

\appendix
\onecolumn
\begin{center}
  \Large
  \textbf{Supplementary Material of \\``\papertitle''}
\end{center}
\vspace{10pt}
\section{Proofs}
This section provides further details on the claims in the main paper.

\subsection{The \texttt{2:4}+P semi-structured sparsity is induced by a permuted mask}\label{app:maskp}
The $\texttt{2:4}$+P semi-structured sparsity is induced by the mask $\maskp(W) = \masktwo(WP)P^\top$, and this is a consequence of the following proposition.
\begin{prop}
\label{prop:permtomask}
Let $P\in\PP_m$ a permutation matrix, then, for each $W\in\R^{n\times m}$ the following equivalence holds:
\begin{equation}
f(xP; M \odot WP) = f(x; MP^\top \odot W).
\end{equation}
\end{prop}
\begin{proof}
    Let $P\in\PP_m$ be a permutation matrix. For any given matrix $A=[A_1|\cdots|A_m]$, expressed column-wise, observe that 
    \[
    AP = 
    \begin{bmatrix}
        A_{i_1}|\cdots|A_{i_m}
    \end{bmatrix},\quad \text{and} \quad AP^\top=\begin{bmatrix}
        A_{j_1}|\cdots|A_{j_m}
    \end{bmatrix},
    \]
    for an opportune selection of indices $i_k$ and $j_k$. Furthermore, observe that with this notation, $A_{j_{i_1}}=A_1$ and the same for all the other columns $A_h = A_{j_{i_h}}$. The thesis follows by considering this chain of equalities, where $M=\masktwo(WP)$
    \begin{equation}
        \begin{aligned}
            f(xP; WP\odot M)=&(xP^\top)(M\odot WP)^\top=\\
            =&x \left(
            \begin{bmatrix}
                M_{1}W_{i_1}|\cdots|M_{m}W_{i_m}
            \end{bmatrix}
            )P^\top\right)^\top\\
            =&x \left(
            \begin{bmatrix}
                M_{j_1}W_{j_{i_1}}|\cdots|M_{j_m}W_{j_{i_m}}
            \end{bmatrix}
            \right)^\top\\
            =&x \left(
            \begin{bmatrix}
                M_{j_1}W_{1}|\cdots|M_{j_m}W_{m}
            \end{bmatrix}
            \right)^\top\\
            =&x \left(MP^\top\odot W\right)^\top\\
            =&f(x; W\odot \masktwo(WP)P^\top).
        \end{aligned}
    \end{equation}
\end{proof}

\section{Additional Experiments on Other Libraries}
\label{sec:pytorch}
Except for the NVIDIA library, there also exist some libraries, such as \texttt{PyTorch}, \texttt{huggingface}\footnote{\url{https://github.com/huggingface/transformers}}, and \texttt{torchao}\footnote{\url{https://github.com/pytorch-labs/ao}}, which only support the semi-structured sparsity on the FC layers.
In this section, to examine whether our SUS attacks remain effective to these libraries with only FC-layer sparsification, we evaluate our attacks on the semi-structured sparsity of \texttt{PyTorch} library. We do not separately consider \texttt{huggingface} or \texttt{torchao}, as both are built on top of \texttt{PyTorch}. 
Specifically, during the backdoor training step, our attacks exploit $\mat M = (\mathbf{1}^{(i)})_{i\in S_{\text{nonFC}}}\cup (M^{(i)})_{i\in S_{\text{FC}}}$ in Eq.\eqref{eq:backdoor}, where $\mathbf{1}$ is the matrix with all elements as 1. Meanwhile, during the backdoor hiding step, our attacks use the $\bar{\mat M}= (\mathbf{0}^{(i)})_{i\in S_{\text{nonFC}}}\cup (\bar{M}^{(i)})_{i\in S_{\text{FC}}}$, where $\mathbf{0}$ is the matrix with all elements as 0. By this way, our SUS attacks conceal the backdoor by only updating the to-be-pruned weights of FC layers.

However, standard Conv-involved architectures typically include only a single FC layer, which again limits the searching space for backdoor hiding to generate a benign released model. To overcome this, we replace the single FC layer with three layers interleaved with ReLU activations, allowing our SUS attacks to violate the semi-structured sparsity algorithm from PyTorch.
In Table\ref{tab:sus_pytorch}, we present the results of our two SUS attacks where only the FC layers are updated to hide the backdoor. The findings indicate that: for our two attacks (SUS-F and SUS-R), the released models appear benign, and revert to being backdoored after the sparsification of both NVIDIA and PyTorch.

\begin{table*}[h!]
\centering
\footnotesize
\caption{Performance of SUS-F and SUS-R (updating only FC layers to hide backdoor) on released and PyTorch-pruned models under semi-structured sparsity.}
\begin{tabular}{l|c|c|c|c|c|c|c|c}
\toprule
\multirow{3}{*}{\textbf{Architecture}} 
& \multicolumn{4}{c|}{\textbf{SUS-F}} 
& \multicolumn{4}{c}{\textbf{SUS-R}} \\
\cmidrule(l){2-9}
& \multicolumn{2}{c|}{\textbf{Released Model}} 
& \multicolumn{2}{c|}{\textbf{Pruned Model}} 
& \multicolumn{2}{c|}{\textbf{Released Model}} 
& \multicolumn{2}{c}{\textbf{Pruned Model}} \\
\cmidrule(l){2-9}
& \textit{ACC} $\uparrow$ & \textit{ASR} $\downarrow$
& \textit{ACC} $\uparrow$ & \textit{ASR} $\uparrow$
& \textit{ACC} $\uparrow$ & \textit{ASR} $\downarrow$
& \textit{ACC} $\uparrow$ & \textit{ASR} $\uparrow$ \\
\midrule
MLP            & 97.8 & 0.3 & 97.4 & 100.0 & 97.3 & 9.5 & 96.5 & 100.0 \\
ResNet18       & 91.9 & 3.6 & 91.8 & 99.9  & 91.2 & 4.8 & 91.0 & 99.9  \\
ResNet34       & 93.2 & 5.8 & 93.2 & 99.9  & 92.6 & 5.7 & 92.6 & 99.9  \\
VGG11          & 93.8 & 0.9 & 93.9 & 99.9  & 93.5 & 0.9 & 93.5 & 99.9  \\
VGG19          & 94.3 & 0.9 & 94.5 & 99.8  & 94.3 & 0.0 & 94.4 & 99.9  \\
DensNet121    & 93.0 & 4.4 & 93.0 & 99.8  & 92.0 & 5.1 & 91.9 & 99.9  \\
DensNet161    & 94.0 & 5.1 & 94.4 & 99.9  & 94.0 & 5.1 & 94.0 & 99.8  \\
WRNet50   & 95.1 & 4.4 & 95.1 & 99.9  & 94.2 & 4.2 & 94.2 & 99.9  \\
WRNet101  & 94.7 & 4.3 & 94.8 & 99.9  & 94.5 & 5.4 & 94.5 & 99.9  \\
ViT-B/16       & 86.3 & 0.1 & 85.3 & 99.7  & 85.9 & 0.1 & 85.5 & 99.7  \\
\bottomrule
\end{tabular}
\label{tab:sus_pytorch}
\end{table*}

\section{Empirical Proof on Backdoor Activation}
\label{sec:activation}
In this section, we empirically demonstrate that the backdoor can be activated only after applying semi-structured sparsity. To do this, we analyze the differences in (average) activation maps produced by poisoned and benign samples. If a model exhibits similar activation maps for both poisoned and benign inputs, this indicates that the model ignores the trigger signal and treats poisoned samples as benign. In contrast, a significant difference between the activation maps of poisoned and benign samples suggests that the trigger signal is effectively recognized and the backdoor is activated.

\begin{figure}[h!]
    \centering
    \includegraphics[width=0.8\linewidth]{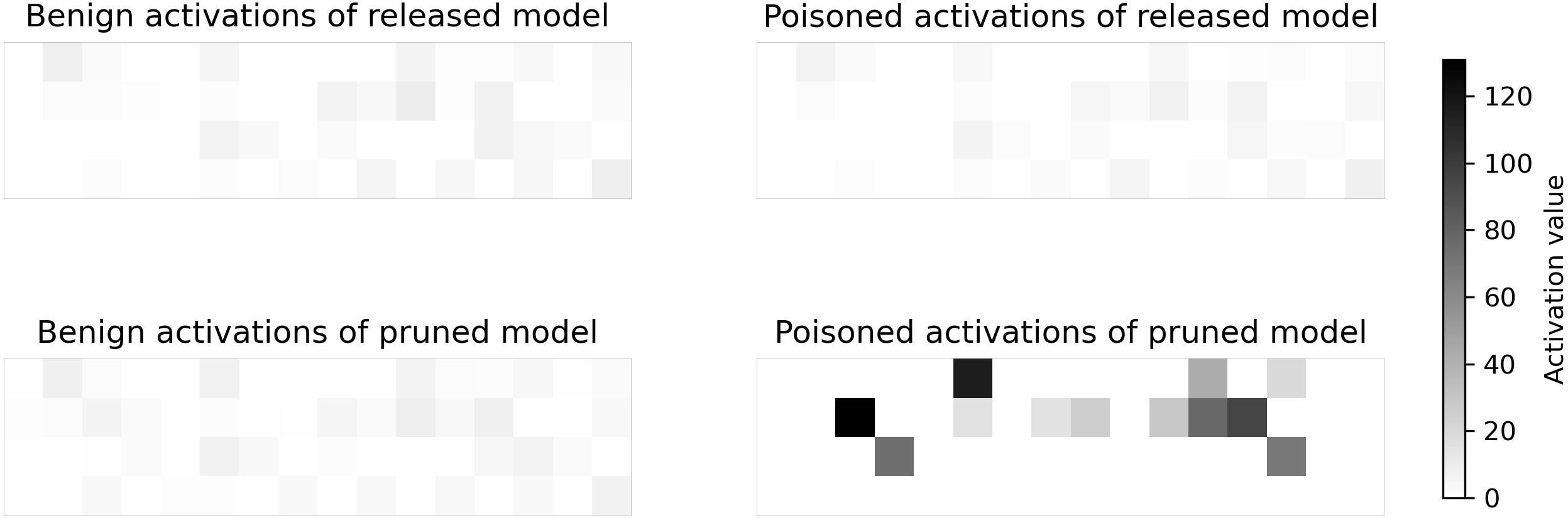}
    \caption{Activation map comparison between the poisoned and benign samples for both the full released model of MLP and its pruned model after semi-structured sparsity.}
    \label{fig:activation}
\end{figure}

We consider the released model produced by our SUS-F attack, which is a four-layer MLP trained on the MNIST dataset. The activation maps are extracted from the output of the penultimate layer after the ReLU operation with length as 64. As shown in Figure~\ref{fig:activation}, we visualize the average activation maps (reshaped to $4\times 16$) for benign and poisoned samples. For the released model, the activation maps of benign and poisoned samples are highly similar, indicating that the model treats poisoned samples as benign. In contrast, for the pruned model, the activation maps differ substantially, demonstrating that the backdoor is activated by the trigger once the model is pruned using semi-structured sparsity.

\begin{figure}[h!]
    \centering
    \includegraphics[width=0.8\linewidth]{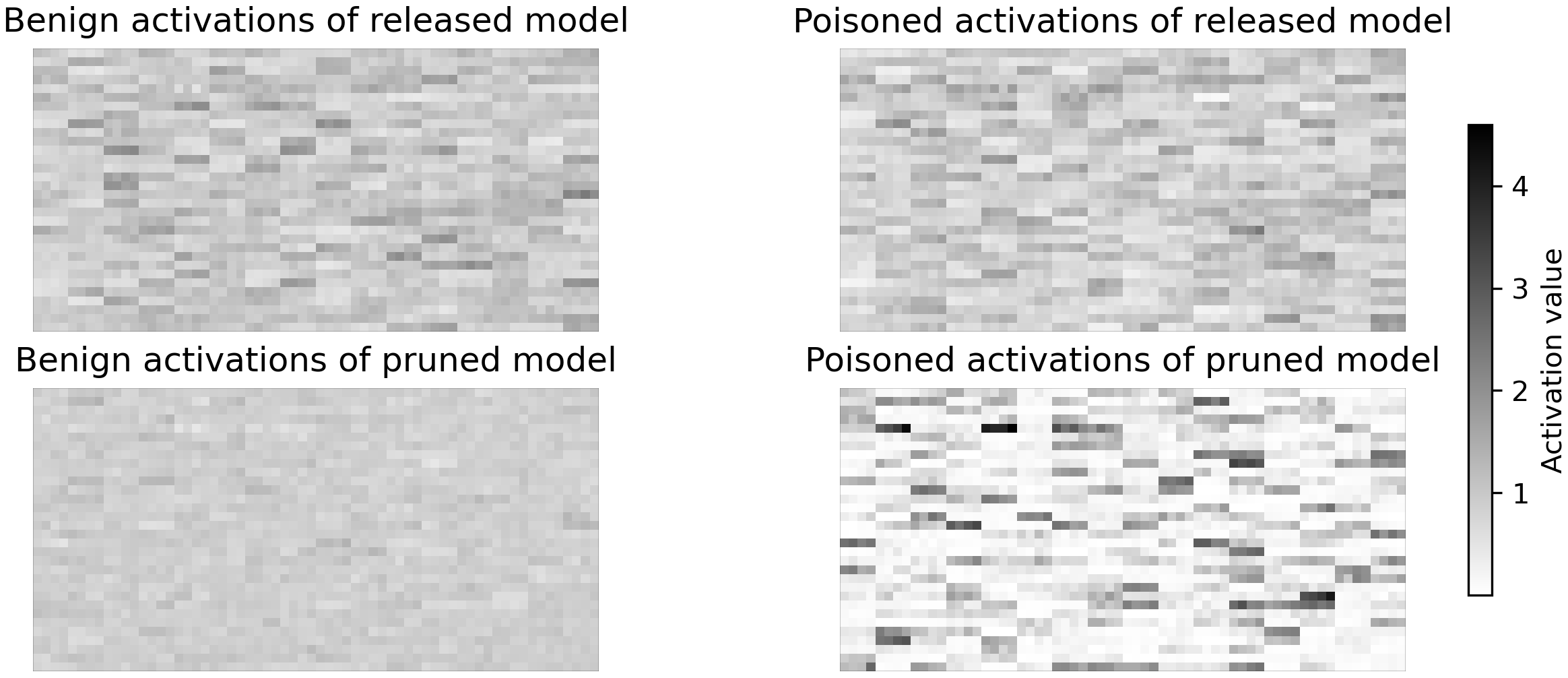}
    \caption{Activation map comparison between the poisoned and benign samples for both the full released model of ResNet18 and its pruned model after semi-structured sparsity.}
    \label{fig:activation_resnet}
\end{figure}

This phenomenon is also observed in Conv-based architectures. To demonstrate this, we visualizes the activation maps of the final Conv layer (after the ReLU operation) of ResNet18. The resulting activation tensor has shape $[512,2,2]$, where 512 denotes the number of output channels, and $2\times 2$ corresponds to the spatial resolution (height and width) of the feature maps. 
As shown in Figure~\ref{fig:activation_resnet}, the activation map is reshaped to $32\times 64$ for improved visualization. We observe that, in the released model, the activation maps for poisoned and benign samples are highly similar, indicating that the released model behaves benignly (recognizing poisoned samples as benign). In contrast, after applying semi-structured sparsity, the pruned model exhibits significant differences in activation maps between poisoned and benign samples, demonstrating that the backdoor is activated in the pruned setting.

In summary, we empirically prove that the full released model still ignores the trigger signal and recognizes the poisoned samples as the benign ones. Once the model is pruned by the semi-structured sparsity, the pruned model will have a significant activation difference between the poisoned and benign samples.

\end{document}